\documentclass[aps,prl,reprint,showpacs,groupedaddress]{revtex4-1}

\usepackage{bm}
\usepackage{wasysym}
\usepackage{amssymb}
\usepackage[cp1251]{inputenc}
\usepackage[dvips]{graphicx}
\usepackage{hyperref}
\begin{document}

\title{Bernstein modes and giant microwave response of two-dimensional electron system\\}

\author{V.A. Volkov}
\email{Volkov.V.A@gmail.com}
\affiliation{Kotelnikov Institute of Radio-engineering and Electronics of RAS, Mokhovaya 11-7, Moscow 125009, Russia\\
}
\affiliation{Moscow Institute of Physics and Technology, Institutskii per. 9, Dolgoprudny, Moscow region 141700, Russia\\}

\author{A.A. Zabolotnykh}
\affiliation{Kotelnikov Institute of Radio-engineering and Electronics of RAS, Mokhovaya 11-7, Moscow 125009, Russia\\
}
\affiliation{Moscow Institute of Physics and Technology, Institutskii per. 9, Dolgoprudny, Moscow region 141700, Russia\\}

\date{\today}

\begin{abstract}
We report on a contribution to the microwave response of a two-dimensional electron system in a magnetic field which originates from excitation of virtual Bernstein modes. These collective modes emerge as a result of interaction between the usual magnetoplasmon mode and cyclotron resonance harmonics. The electrons are found to experience a strongly enhanced radiation field when its frequency falls in a gap of the Bernstein modes spectrum. This field can give rise to nonlinear effects, one of which, the parametric cyclotron resonance, is discussed. We argue that this resonance leads to a plasma instability in the ultraclean system. The instability-induced heating is responsible for the giant photoresistivity spike recently observed in the vicinity of the second cyclotron resonance harmonic. 

\end{abstract}

\maketitle

Plasma oscillations are well known; their study began about a century ago by Langmuir and Tonks. These oscillations are mostly investigated in two different types of systems: low-density nondegenerate gas plasma \cite{Stix} and the degenerate plasma of solids \cite{Platzman}. In the latter case plasma excitations are often referred to as plasmons. A characteristic feature of the solid state systems is that the motion of particles can be easily restricted in one or more directions and thus low-dimensional systems can be created. Properties of plasmons in these systems differ dramatically from those in the three-dimensional (3D) systems. For instance, the plasmon dispersion law in a two-dimensional (2D) electron system (ES) can be written as $\omega_p^{(2)}(q)=\sqrt{2\pi n_s e^2q/\varkappa m}$, where $n_s$ is 2D electron concentration, $\varkappa$ is the surrounding dielectric constant, $m$ is electron effective mass, and $q$ is the 2D wave vector of the plasmon. The spectrum of 2D plasmons is gapless and strongly depends on $q$ in contrast to the dispersionless spectrum of the 3D plasmon $\omega_p^{(3)}(q)=\sqrt{4\pi n_{3D} e^2/\varkappa m}$, where $n_{3D}$ is the electron concentration of 3DES.

Recently an ultrastrong radiation-plasmon coupling in 2DES in magnetic field has been observed \cite{Muravev, Scalari}.
Plasma oscillation in the perpendicular magnetic field $B$ is called the upper hybrid mode in gas plasma or the magnetoplasmon mode in solid state plasma. The dispersion relation of the excitation is as follows:
\begin{equation}
 \label{mp_disp}	
 \omega_{mp}=\sqrt{\omega_c^2+\left(\omega_p^{(i)}\right)^2},
\end{equation} 
where $\omega_c=eB/mc$ is the electron cyclotron frequency, and $\omega_p^{(i)}$ is the plasmon frequency in a 3D ($i=3$) or 2D ($i=2$) system in the absence of a magnetic field.

\begin{figure}
			\includegraphics[width=8.5cm]{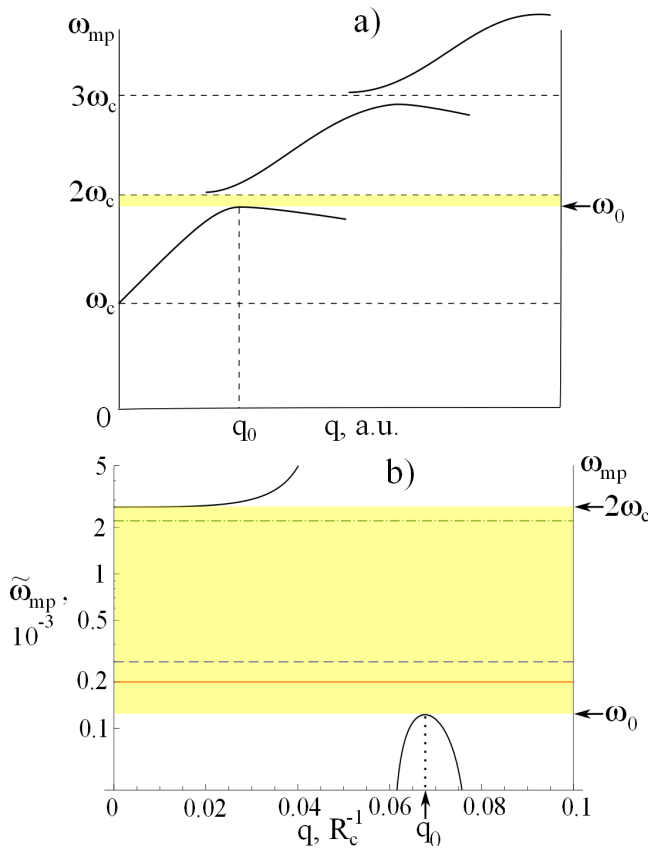}
			\caption{ \label{Fig:b_modes}
			The Bernstein modes in spectra of 2D magnetoplasma excitations. Radiation field with frequency $\Omega$ is strongly modified when $\Omega$ falls in one of frequency gaps (the lowest gap near $2\omega_c$ is shadowed).
			(a) Schematic picture of the Bernstein modes;  $\omega_0$ and $q_0$ correspond to the bottom of the lowest gap.  
			(b) The Bernstein modes dispersion near the $2\omega_c$,  $\tilde{\omega}_{mp}=\omega_{mp}/\omega_c-1.9973$, logarithmic scale. Different lines correspond to the different values of $\Omega$: $\Omega/\omega_c=1.9995$ (dash-dotted line), $\Omega/\omega_c=1.99757$ (dashed line), $\Omega/\omega_c=1.9975$ (solid line). Parameters of 2DES are given in the text. 
		}
\end{figure}

Plasma oscillations determined by Eq.~(\ref{mp_disp}) can interact with cyclotron resonance harmonics due to finite value of $qR_c$, where $R_c=v_F/\omega_c$ is the electron cyclotron radius, and $v_F$ is the Fermi velocity (in the case of degenerate plasma). This interaction splits mode (\ref{mp_disp}) into the so-called Bernstein modes. In 2DES these modes, see Fig.~\ref{Fig:b_modes}a, are separated from each other by gaps situated near $N\omega_c$, $N=2,3..$. Scattering of electrons by impurities can smear these gaps \cite{Chaplik}.
The Bernstein modes are familiar in the physics of gas plasma \cite{I.B.} and 3D solid state plasma \cite{Platzman}. They were also studied both theoretically \cite{Chiu,Horing,Glasser} and experimentally \cite{Kotthaus, Batke, Bangert, Richards, Holland} in 2DESs under different conditions including the quantum Hall regime \cite{Kukushkin_Smet,Kukushkin}.  
 
\begin{figure}
			\includegraphics[width=8.5cm]{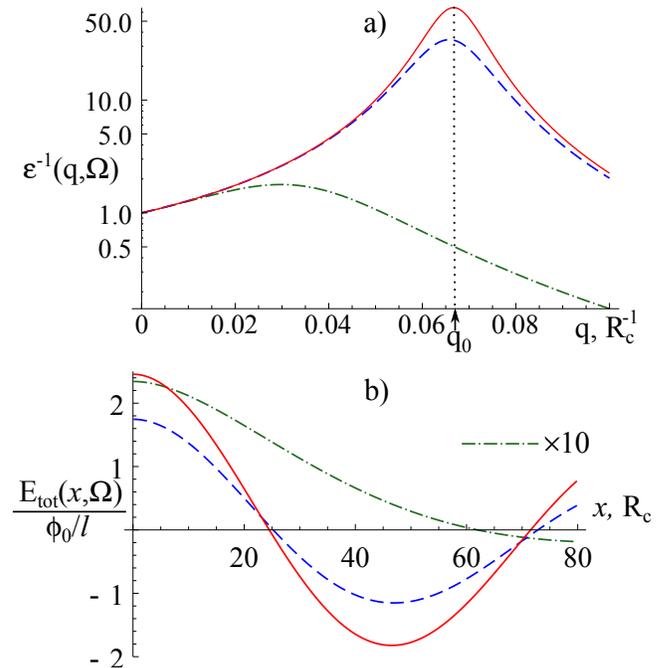}
			\caption{ \label{Fig:inv_eps} (a) Inverse dielectric function $\varepsilon^{-1}(q,\Omega)$ describing the enhancement of the radiation electric field due to the excitation of the virtual Bernstein modes. The radiation frequency $\Omega$ lies in the Bernstein gap ($2\omega_c>\Omega>\omega_0$).
			(b) The amplitude of total electric field in $x$ space. The model function of unscreened field $E_{\Omega}(x)$ induced by irradiated metal contact is use;, see details in the text.
			 Different lines correspond to different values of $\Omega$; see the legend of Fig. \ref{Fig:b_modes}.
}
\end{figure}

Consider the influence of the Bernstein modes on the screening of incident radiation by magnetoplasmons in 2DES. If a wave vector of radiation $q$ is nonzero and the radiation frequency $\Omega$ lies in one of the gaps, see Fig.~\ref{Fig:b_modes}(b), then real magnetoplasma modes are not excited but electric field of radiation can be strongly modified due to a polarization of 2DES. The effect can be described in plasmon terms as an excitation of virtual Bernstein modes with the same frequency and wave vector.
If the frequency $\Omega$ occurs in the Bernstein gap and goes to the frequency of single-particle excitations $N\omega_c$, i.e., to the top of the gap, then the longitudinal dielectric function $\varepsilon(q,\omega)$ becomes infinite at any $q$ and the total electric field in the system vanishes. It is the regime of the usual screening; see the dash-dotted lines in Figs.~\ref{Fig:b_modes} and \ref{Fig:inv_eps}. 
But if $\Omega$ and $q$ are close to the bottom of the gap $(\omega_0,q_0)$, where $\varepsilon(q_0,\omega_0)=0$, then the corresponding Fourier component of the radiation field is amplified due to a factor $1/\varepsilon(q,\omega)$, Fig.~\ref{Fig:inv_eps}(a). As a result, the total electric field is enhanced and even becomes oscillating in space; see dashed and solid lines in Figs.~\ref{Fig:b_modes} and \ref{Fig:inv_eps}.

The enhanced response due to the Bernstein modes excitation opens a new area of nonlinear effects studies in 2D magnetoplasma. These phenomena are expected to be revealed in high-mobility systems in which a smearing of the Bernstein gaps is small.
If the magnetic field is weak ($0.1-1$ T) then typical frequencies of the Bernstein modes for GaAs/AlGaAs 2D structures lie in the microwave (MW) range and one can expect these effects to appear in experiments on MW-irradiated 2DESs.

Below we consider one of the nonlinear effects, cyclotron parametric resonance. The parametric resonance in 2DESs can be understood using an analogy with a simple pendulum whose length varies periodically with frequency $\Omega$. The fundamental mode of parametric resonance develops at $\Omega$ equal to the double eigenfrequency of the pendulum. In a 2DES, $\omega_c$ acts as the eigenfrequency and the fundamental cyclotron parametric resonance mode is excited at $\Omega\approx 2\omega_c$.

We suppose that a plasma instability due to this cyclotron parametric resonance was observed in recent experiments on the MW photoresistance of an ultraclean 2DES in a magnetic field \cite{Dai_Du,Dai_Stone,Hatke_1,Hatke_2}. In these papers there was reported a giant photoresistance spike that appears at MW radiation frequency $\Omega$ close to $2\omega_c$. The spike is much higher and narrower than ordinary photoresistance maxima observed in high-mobility structures in the regime of MW-induced resistance oscillations \cite{Zudov, Ye, review}. It is important that all 2DESs featuring the spike exhibit a temperature-sensitive giant negative magnetoresistance (GNMR); that is why instability-induced heating of 2DESs leads to the photoresistivity spike.
An attempt to explain the spike was made in Ref.~\cite{Inarrea}, but its origin still remains unclear. Note also that parametric resonances under different conditions were studied theoretically in Refs.~\cite{par_Orgad, par_Mikh, par_Joas, par_Weick}. 
 
This Rapid Communication is organized as follows. We evaluate first the total electric field of radiation which is enhanced by the Bernstein modes. Then we develop a hydrodynamic theory of cyclotron parametric resonance taking into account the enhanced radiation field. Finally, we find the conditions of the plasma instability to occur in 2DESs due to excitation of the fundamental mode of the parametric resonance.

We consider a MW-irradiated 2DES positioned in plane $z=0$ and placed into the perpendicular magnetic field $\bm{B}=(0,0,B)$. As a rule, the wavelength of incident radiation is greater than a typical size of the sample. So, the MW electric field is inhomogeneous on the cyclotron radius scale, for example, because of the metal contacts to the 2DES, which significantly modify the MW radiation field~\cite{Mikh}.
This electric field is defined as $\bm{E}_0(\bm{r},t)=\bm{E}_{\Omega}(\bm{r})\cos\Omega t$, with an amplitude $\bm{E}_{\Omega}(\bm{r})$ dependent on the coordinates $\bm{r}=(x,y)$. For simplicity we consider electric field $\bm{E}_{\Omega}(\bm{r})$ directed along the $x$ axis and assume that it depends only on the $x$ coordinate.
The MW electric field $E_{\Omega}(x)$ and its Fourier component $E_{\Omega}(q)$ are screened by the Bernstein modes of 2D magnetoplasma. We use linear response theory to evaluate the total electric field $E_{tot}(x,t)=E_{tot}(x,\Omega)\cos\Omega t$ and the random phase approximation to find the dielectric function of 2DES $\varepsilon(q,\omega)$:
\begin{equation}
\label{screening}
	E_{tot}(x,\Omega)=\varint^{+\infty}_{-\infty}\frac{dq}{2\pi} e^{iqx}\frac{E_{\Omega}(q)}{\varepsilon(q,\Omega)}.
\end{equation}

In the collisionless limit, at $q\ll k_F$, $\hbar\omega_c\ll 2\pi^2k_B T\ll E_F$, where $\hbar k_F$, $E_F$ are Fermi momentum and energy, $T$ is temperature, and the function $\varepsilon(q,\omega)$ is determined as \cite{Chiu,Greene,Volkov}
\begin{equation}
\label{diel_per}
	\varepsilon(q,\omega)=1+\frac{2m}{\pi\hbar^2}V_{ee}(q)\sum_{n=1}^{\infty} \frac{ n^2\omega_c^2 J_n^2(qR_c)}{n^2\omega_c^2-\omega^2-i0sgn\omega},
\end{equation}
where $V_{ee}(q)=2\pi e^2/\varkappa |q|$ is the Fourier component of the 2D Coulomb potential, and $J_n (qR_c)$ is the $n$th-order Bessel function of the first kind. 

Let us estimate the frequency of the bottom of the gap $\omega_0$, its position $q_0$, and the gap width $\Delta_0\equiv 2\omega_c -\omega_0$. In the lowest order in the small parameter $qR_c$ we obtain
	$q_0R_c=4.5 a_B/R_c$,
where $a_B=\hbar^2\varkappa/me^2$ is the effective Bohr radius. For $\omega_c/2\pi=100$~GHz and typical parameters of GaAs quantum wells ($\varkappa=7$, $m=0.067m_0$, $n_s=3\times 10^{11}$ cm$^{-2}$) we obtain $a_B\approx 5.5$ nm, $B\approx 0.24$~T, $R_c\approx 0.37$ $\mu$m, and $q_0R_c\approx 0.067\ll 1$.
In the same approximation the gap width is estimated as
\begin{eqnarray}
	\nonumber
	\frac{\Delta_0}{2\pi}\approx \frac{11.4}{2\pi}\frac{a_B^2}{R_c^2}\omega_c
	\approx 0.24\left[\frac{\omega_c/2\pi}{100\,GHz}\right]^3[GHz].
	\nonumber
\end{eqnarray}
From Eqs. (\ref{screening}), (\ref{diel_per}) we are able to find an asymptotics of the total electric field $E_{tot}(x,\Omega)$: $E_{tot}(x,\Omega)\simeq E_{\Omega}(q_0)\cos q_0x/\varepsilon(q_0, \Omega)$, when $\Omega\to\omega_0+0$. The total radiation field is enhanced strongly due to $\varepsilon(q_0,\Omega)\to 0$ at this limit. We have computed $E_{tot}(x,\Omega)$ for the above-mentioned experimental parameters, Fig.~\ref{Fig:inv_eps}(b).

Consider the parametric resonance of electrons induced by the enhanced radiation field. 
We describe the motion of electrons by the Euler equation for the hydrodynamic velocity $\bm{V}=\bm{V}(x,t)$:
\begin{equation}
\label{hydro}
\partial_t\bm{V}+\frac{\bm{V}}{\tau}+(\bm{V},\bm{\nabla})\bm{V}  = \frac{e}{m}\bm{E}_{tot}(x,t)+\frac{e}{mc}[\bm{V},\bm{B}],
\end{equation}
where $\tau$ is the phenomenological relaxation time. 
We assume than the mobility in the 2DES is high and conditions $\Omega\tau\gg 1$, $\omega_c\tau\gg 1$ take place.
The nonlinear term $(\bm{V},\bm{\nabla}) \bm{V}$ in~(\ref{hydro}) plays a central role in our approach. It can be interpreted as a nonlinear, local, and instantaneous Doppler shift of the frequency of excitations described by Eq.~(\ref{hydro}).

Solutions of Eq.~(\ref{hydro}) for the forced oscillations of velocity $\bm{V}_0=(V_{0x},V_{0y})$ can be written in the linear approximation as $V_{0x}(x,t)=V_{sx}(x)\sin\Omega t$, $V_{0y}(x,t)=V_{cy}(x)\cos\Omega t$, where 
\begin{equation}
	\label{vel_lin}
	V_{sx}(x)= \frac{eE_{tot}(x,\Omega)\Omega}{m(\Omega^2-\omega_c^2)},
	V_{cy}(x)= V_{sx}(x)\frac{\omega_c}{\Omega}.
\end{equation}

To find a nonlinear correction $\delta\bm{V}(x,t)$ to the velocity we substitute $\bm{V}(x,t)=\bm{V}_0(x,t)+\delta\bm {V}(x,t)$ into Eq.~(\ref{hydro}) and derive the following exact set of equations for $\delta \bm{V}=(\delta V_x,\delta V_y)$:
\begin{equation}
\label{param}
	\left\{ 
		\begin{array}{lcr}
			(\partial_t+1/\tau)\delta V_x+ V_{0x}'\delta V_x +(V_{0x}+ \delta V_x)\delta V_x' \\
			\qquad-\omega_c\delta V_y=-V_{0x}V_{0x}'\\
			\\
			(\partial_t+1/\tau)\delta V_y+ V_{0y}'\delta V_x +(V_{0x}+ \delta V_x)\delta V_y'\\ \qquad+\omega_c\delta V_x= -V_{0x}V_{0y}'
			 \end{array},
	\right.
\end{equation}
where the prime denotes the derivative with respect to $x$.

The third and the forth terms on the left-hand side of Eqs.~(\ref{param}) as well as the right-hand side terms in these equations stem from the nonlinear term $(\bm{V},\bm{\nabla})\bm{V}$ in Eq.~(\ref{hydro}). The coefficients in the third and the fourth terms on the left-hand side of Eqs.~(\ref{param}) periodically depend on time and cause the parametric resonance in the 2DES.
The variable $\delta V_y(x,t)$ can be excluded from Eqs.~(\ref{param}) and the obtained equation is linearized with respect to $\delta V_x$. We do not present an explicit form of this cumbersome equation.

Following the standard procedure of the parametric resonance theory \cite{Landau}, a solution for the fundamental mode at $\Omega\approx 2\omega_c$ can be written in the two-wave approximation as:
\begin{equation}
\label{sol}
	\delta V_x=e^{s_0t} \left[A(x)\cos\frac{\Omega t}{2}+ B(x)\sin\frac{\Omega t}{2}\right]V_{sx}^{-1}(x),
\end{equation}
where $s_0$ is the amplification coefficient, $|s_0|\ll\Omega$. The instability occurs at $s_0>0$.

We substitute Eq.~(\ref{sol}) into the linearized equation for $\delta V_x(x,t)$ to obtain the set of the linear equations for the coefficients $A(x)$, $B(x)$. In this derivation, we neglect the terms containing high order frequency harmonics as well as the nonlinear terms with respect to the electric field amplitude.
Neglecting also the difference between $\Omega$ and $2\omega_c$ where it is possible,
we arrive to the system of homogeneous equations for $A(x)$ and $B(x)$. 
Then we exclude function $B(x)$ and simultaneously omit the terms small by the parameter $s/\Omega\ll 1$, where $s=s_0+1/\tau$. For simplicity we also consider the ''clean limit,'' assuming that the Bernstein gap smearing is small: $s^2\ll \Delta^2$, where $\Delta\equiv2\omega_c-\Omega$. With these assumptions the equation for $A(x)$ takes the form
\begin{equation}
\label{squared}
	\left(-\partial_x^2 - \left(\frac{\Omega^2}{4s^2}\right)
	\frac{(\frac{3}{4}V_{sx}')^2-\Delta^2}{V_{sx}^2}  \right)  A(x)=0.
\end{equation}

Equation~(\ref{squared}) is formally equivalent to the Schr\"odinger-type equation with an effective potential energy  
\begin{equation}
	\label{pot_energy}
	U(x)=
-\left(\frac{\Omega^2}{4s^2}\right)
\frac{(\frac{3}{4}V_{sx}')^2-\Delta^2}{V_{sx}^2}
\end{equation}
and zero effective energy. For solutions of Eq.~(\ref{squared}) to exist, the potential energy $U(x)$ should be attractive. 
\begin{figure}
			\includegraphics[width=8.5cm]{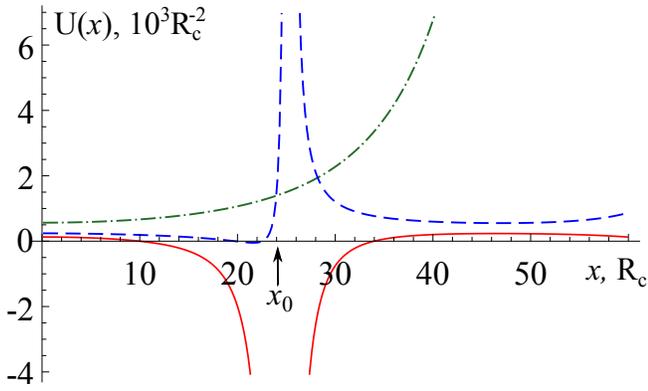}
			\caption{ \label{Fig:U} The effective potential energy (\ref{pot_energy}) in the Schr\"odinger-type Eq.~(\ref{squared}). Different lines correspond to different values of $\Omega$; see the legends of Figs. \ref{Fig:b_modes}, \ref{Fig:inv_eps}. Solid line formally corresponds to the quantum-mechanical phenomena, named ''fall to the center.'' As a result, a plasma instability develops in the system near $x=x_0\sim\pi/2q_0$.
}
\end{figure}
The computed $U(x)$ for different $\Omega$ are presented in Fig.~\ref{Fig:U}. The parameters of the 2DES were as listed above. We also use the model function of electric field $E_{\Omega}(x)=\phi_0/\sqrt{l^2+x^2}$, where $\phi_0/l$ is the characteristic value of the field inhomogeneity. We assume $\phi_0=1$ mV, $l=R_c$ and $\Omega/s=700$. The function $E_{\Omega}(x)$ describes the edge field of irradiated metal contact \cite{Mikh_2}. The total field $E_{tot}(x,\Omega)$ is weakly sensitive to the length $l$, because $E_{\Omega}(q)\approx 2\phi_0 \ln(1.12/ql)$ at $ql\ll 1$. 

We see in Figs.~\ref{Fig:inv_eps} and \ref{Fig:U} that as $\Omega$ approaches the bottom of the gap $\omega_0$, the amplitude of screened field $E_{tot}(x,\Omega)$ [and $V_{sx}(x)$] increases and becomes oscillating. Note that $E_{tot}(x,\Omega)$ changes its sign, for the first time at $x=x_0\sim\pi/2q_0$. Near this point $V_{sx}(x)= V_{sx}'(x_0)(x-x_0)$ and  
$U(x)$ is attractive, if $V_{sx}'(x_0)\gtrsim 4\Delta/3$. Moreover the potential energy has asymptotic behavior $-C/(x-x_0)^2$, where $C\gg 1$ in the limit $\Omega\to\omega_0$. Such an asymptotics of potential energy corresponds to the so-called ''fall to the center'' in quantum mechanics \cite{Landau_3}. In this case solutions of Eq.~(\ref{squared}) always exist and a local plasma instability develops near $x=x_0$. 
In turn, it leads to a heating of the 2DES which destroys the temperature-sensitive GNMR. Therefore the photoresistivity peak arises at $\Omega$ close to ($2\omega_c-\Delta_0$).

We now discuss other conditions of the plasma instability to appear. The condition $\tau^*\Delta_0\sim 1$ defines the minimal frequency of the MW radiation. Here $\tau^*$ is an effective relaxation time that defines the Bernstein gap smearing. For a 2DES with the parameters listed above and $\mu=e\tau/m= 3\times 10^7$ cm$^2$/(Vs) we are able to estimate this minimal frequency assuming $\tau^* \apprge \tau$ as     
\begin{eqnarray}
\frac{\Omega_{min}}{2\pi}=\frac{1}{2\pi}\sqrt[3]{\frac{8}{11.4}\left(\frac{v_F}{a_B}\right)^2 \frac{1}{\tau^*}} \apprle 160 \  GHz.\nonumber
\end{eqnarray}

One can also estimate the minimal effective electron mobility needed for the instability to take place at typical frequency $\Omega/2\pi=100$~GHz:
$\mu^*_{min}=e\tau^*/m\approx 13\times 10^7$ cm$^2/$(Vs).
One should point out that the instability due to parametric resonance develops at the distances from the contact of the order $\pi/2q_0\sim 24R_c\approx 9\,\mu m$. Certainly, this distance should be smaller than the sample's width. 

Discuss also the applicability of the semiclassical approximation for dielectric function $\varepsilon(q,\omega)$. Equation (\ref{diel_per}) is formally valid if the parameter $\hbar\omega_c/ 2\pi^2k_B T$ is small, but this parameter is of the order of unity in the experiments \cite{Dai_Du,Dai_Stone,Hatke_1,Hatke_2} . Let us take into account that the  Shubnikov-de Haas oscillations disappear in a field $B\approx 0.1$ T even at low temperatures \cite{Dai_Du,Hatke_1}. This effect is likely due to large-scale electron-density fluctuations which can be described by introducing an effective temperature of the system $T^*\approx \hbar\omega_c/k_B=2$~K at $B=0.1$~T \cite{Volkov}. The condition $2\pi^2k_BT^*\gg \hbar\omega_c$ is satisfied and it allows us to use Eq. \ref{diel_per} to explain experimental data \cite{Dai_Du,Dai_Stone,Hatke_1,Hatke_2}.   

In summary, we consider the mechanism of MW field enhancement due to excitation of the virtual Bernstein modes of a 2D magnetoplasma. The mechanism is realized only in ultraclean 2DESs in which the Bernstein gap is larger than the gap smearing due to scattering. The enhanced field leads to the appearance of the cyclotron parametric resonance, which takes place when $\Omega$ is close to $2\omega_c$. We argue that the excitation of the fundamental mode of the cyclotron parametric resonance is the reason for the giant MW response recently discovered in ultraclean 2DESs near $2\omega_c$. We show that fluctuations of hydrodynamic velocity obey the Schr\"odinger-type equation. An effective potential energy in this equation corresponds to the ''fall to the center'' if $\Omega$ is close to the bottom of the Bernstein gap near $2\omega_c$. As a result the plasma instability occurs. In turn, it leads to electron heating and the giant photoresistance spike observed in experiments \cite{Dai_Du,Dai_Stone,Hatke_1,Hatke_2}.

\begin{acknowledgments}
We thank I. V. Kukushkin and M. Zudov for a critical reading of the manuscript and G. R. Aizin for numerous useful discussions. The work was supported by the Russian Foundation for Basic Research (Project No. 14-02-01166) and the Foundation ''Dynasty''.
\end{acknowledgments}

\end{document}